# The Geometry of Crashes
## - A Measure of the Dynamics of Stock Market Crises

*Tanya Araújo[(1)] and Francisco Louçã [(2)]*

Departamento de Economia, ISEG, Technical University of Lisbon
Research Unit on Complexity in Economics (UECE)
[(1)] tanya@iseg.utl.pt    [(2)] flouc@iseg.utl.pt

### Abstract

*This paper investigates the dynamics of stocks in the S&P500 index for the last 30 years. Using a stochastic geometry technique, we investigate the evolution of the market space and define a new measure for that purpose, which is a robust index of the dynamics of the market structure and provides information on the intensity and the sectoral impact of the crises. With this measure, we analyze the effects of some extreme phenomena on the geometry of the market. Nine crashes between 1987 and 2001 are compared by looking at the way they modify the shape of the manifold that describes the S&P500 market space. These crises are identified as (a) structural, (b) general and (c) local.*

*JEL C3, G8*

## 1 Introduction

In 1999, R. Mantegna [1] defined a distance metric based on correlation coefficients between the log-price difference of a pair of market securities. This metric allows for determining a distance between stocks evolving in time in a synchronous fashion. Since the metric was further discussed by Mantegna and Stanley [2] in the book that coined the term "Econophysics", it has been applied in a considerable number of research works ([3]-[11]). The fact that the metric is a properly defined distance gives a meaning to geometric notions in the study of the market. As Mantegna did when the distance was first introduced [1], many papers using the metric follow a topological approach.

Provided that a distance between stocks exists, it is sufficient to form an additional hypothesis on the topological space of the stocks (as for example, choosing the subdominant ultrametric space, which is obtained from the minimal-spanning tree that links the stocks [2]) in order to end up with a connectivity pattern for the stocks. In so doing, one can naturally move away from a situation in which all the stocks were connected to a network of stocks, in which the connectivity pattern was endogenously determined. From the topological point of view, it opens a large set of promising possibilities to explore.

Using Mantegna's metric we followed a different perspective. In a previous contribution [12] we developed a method for the reconstruction of an economic space. By using a stochastic geometry technique, we proved that economic spaces are low-dimensional entities and that this low-dimensionality is caused by the small proportion of systematic information present in correlations among stocks. Using our reconstruction method we found that part of the correlation contribution is virtually indistinguishable from random data.

In the present paper, we investigated the hypothesis that market spaces uniformly contract during crashes along their effective dimensions and concluded that, otherwise, some crashes may act differently on specific directions, causing interesting changes in the shape of the market space. In order to capture that distortion effect, the evolution of the market space is verified as it is reconstructed under a moving window over an interval of 16 days. A structure index is then used to compute the lack of uniformity among the market effective dimensions. As a consequence, we are able to characterize the structures that emerge in relevant historical periods and to identify the economic sectors that are associated to important changes in the leading directions of the evolving market space.

It is empirically observed that both during expansion and normal periods the market tends toward randomness whereas in the disturbed periods its structure is reinforced, not only in the topological sense (as revealed by the clustering measures) but also in the geometrical sense, considering distortions of form. From this observation we propose a new measure of the dynamics of the market structure, which captures that distortion effect in the shape of the market space.

Some other authors also discussed the existence of a dynamic pattern during market's crashes ([10], [14]-[20]). Sornette and his co-authors successfully demonstrated that some dynamic patterns can often be found in preceding events. For several extreme phenomena, they found evidence of incoming instabilities in the precursory patterns of time trajectories of market data (as price, volume and volatility variables). Among their main contributions, there is an issue that appears to be crucial for understanding the behavior of the market: the identification of distinct signature for endogenous and exogenous shocks originating crashes. In particular, they proved a systematic association of large events with positive feedback processes. Later in the paper we shall address that issue while applying our structure index to discriminate distinct processes at work in the S&P500 stock market.

The identification of economic sectors as clusters of stocks with a similar economic dynamics was discussed in references [8] to [11]. In reference [9], Gopikrishnan *et al.* used techniques that are related to the metric we use, although with a different perspective. Diagonalizing the correlation matrix, they have tried to identify particular eigenvectors with the traditional industrial sectors. In our analysis the effective dimensions of a market space may not correspond to economic sectors. We argue that the lack of uniformity among the effective dimensions reveals the existence of a dynamic pattern (which we empirically verify that correspond to crashes). To evaluate the impact of those extreme phenomena in different economic sectors (and the sectoral dynamics among different crashes), we compute the index of market structure for different market spaces, each of them comprising stocks that belong to a specific economic sector.

In sections 2 and 3 the method is explained in detail and it is applied to a set of companies that are or have been in the S&P500 index. In section 4 we discuss the results obtained for specific sectors and the role of those sectors in some important market crashes. Finally, a summary and conclusions are presented.

## 2 Method

The idea is simply stated in the following terms:

1) Pick a representative set of N stocks and their historical data of returns over some time interval.
2) Using an appropriate metric, compute the matrix of distances between the N stocks.
3) From the matrix of distances compute the coordinates for the N stocks in an Euclidean space of dimension $D \leq N-1$.
4) Apply the standard analysis of reduction of the coordinates to the center of mass and compute the eigenvectors of the inertial tensor.
5) Apply the same technique to random data with the same mean and variance.
6) Compare the eigenvalues in (4) with those in (5) and identify the directions for which the eigenvalues are significantly different as being the market characteristic dimensions.
7) From the eigenvalues of order smaller than the number of characteristic dimensions, compute the difference between eigenvalues in (4) with those in (5). The normalized sum of those differences is the index $\boldsymbol{S}$, which measures the evolution of the distortion effect in the shape of the market space.

For both random and actual data, the sorted eigenvalues, from large to small, decrease with their order. In the random case, the amount of decrease is linear in the order number, proving that the directions are being extracted from a spherical configuration. The display of a uniform and smooth decrease in the values of the sorted eigenvalues is characteristic of random cases and is also experimentally observed when the market space is built from historical data corresponding to a period of *business as usual*.

Considering for the lack of uniformity among the market effective dimensions we are able to characterize the extent to which crashes act differently on specific directions, causing changes in the shape of the market space. Looking for relevant distortions in the shape of the S&P500 market space through the last 30 years, we found that amongst the highest values of the index are those computed in some important dates, as 19[th] October 1987, 11[th] September 2001 and 27[th] October 1997.

In addition to the geometrical analysis of the whole S&P500 market space, our measure is applied to sets of stocks that belong to specific economic sectors. Results show that some crashes act differently on specific sectors and that the deviation from random behavior may be limited to a few days after the day of the crash and also to a small number of sector-oriented groups of stocks. Accordingly to these characteristics, crises are identified as (a) structural, (b) general and (c) local.

## 3. Measures

From the returns $r(k)$ for each stock

$$r(k) = \log(p_t(k)) - \log(p_{t-1}(k)) \tag{1}$$

a normalized vector

$$\vec{\rho}(k) = \frac{\vec{r}(k) - \langle \vec{r}(k) \rangle}{\sqrt{n\left(\langle \vec{r}^{\,2}(k) \rangle - \langle \vec{r}(k) \rangle^2\right)}} \tag{2}$$

is defined, where $n$ is the number of components (number of time labels) in the vector $\vec{\rho}(k)$. With this vector one defines the distance between the stocks $k$ and $l$ by the Euclidian distance of the normalized vectors.

$$d_{ij} = \sqrt{2(1 - C_{ij})} = \left\| \vec{\rho}(k) - \vec{\rho}(l) \right\| \tag{3}$$

as proposed in [1], with $C_{ij}$ being the correlation coefficient of the returns $r(i)$, $r(j)$. The fact that $d_{ij}$ is a properly defined distance gives a meaning to geometric notions and geometric tools in the study of the market.

Given that set of distances between points, the question now is reduced to an embedding problem: one asks what is the smallest manifold that contains the set. If the proportion of systematic information present in correlations between stocks is small, then the corresponding manifold will be a low-dimensional entity. The following stochastic geometry technique was used for this purpose.

### 3.1 The stochastic geometry technique

After the distances ($d_{ij}$) are calculated for the set of $N$ stocks, they are embedded in $\Re^D$, where $D \le N-1$, with coordinates $\{\vec{x}(k)\}$. The center of mass $\vec{R}$ is computed and coordinates reduced to the center of mass.

$$\vec{y}(k) = \vec{x}(k) - \vec{R} \tag{4}$$

and the inertial tensor

$$T_{ij} = \sum_k y_i(k) y_j(k) \tag{5}$$

is diagonalized to obtain the set of normalized eigenvectors $\{\lambda_i, \vec{e_i}\}$. The eigenvectors $\vec{e_i}$ define the characteristic directions of the set of stocks. The characteristic directions correspond to the eigenvalues $(\lambda_i)$ that are clearly different from those obtained from random data. They define a reduced subspace of dimension $d$, which carries the systematic information related to the market correlated structure [12].

## 3.2 Index of the market structure

Since the largest $d$ eigenvalues define the effective dimensionality of the economic space, we compute $S$ as:

$$S_t = \sum_{i=1}^{d} \frac{\lambda_t(i) - \lambda'(i)}{\lambda'(i)} = \sum_{i=1}^{d} \frac{\lambda_t(i)}{\lambda'(i)} - 1 \tag{6}$$

where $\lambda_t(1)$, $\lambda_t(2)$, ..., $\lambda_t(d)$ are the largest $d$ eigenvalues of the market space and $\lambda'(1)$, $\lambda'(2)$, ... $\lambda'(d)$ are the largest $d$ eigenvalues obtained from random data over the same time window and with the same mean and variance.

Vilela Mendes proposed in [13] an index that quantifies the effect of some structure-generating mechanisms in dynamical models, based on the fact that a structure in a collective system acquires a characteristic length larger than that of the individual components of the system. We develop this strategy for the definition of our structure index $S$: as the dynamics of systems develop a structure-generating mechanism, the index $S$ measures the normalized difference between the characteristic length of those structures and the characteristic length of the individual components of the system. This is a geometrical approach to define and to measure emergence.

In portfolio optimization models, when the systematic and unsystematic contributions to the portfolio risk are distinguished, the former is associated to the correlation between stocks (collective structure) and the later to the individual variances of each stock [12]. Consequently, when $S$ is applied to the market space, the eigenvalues obtained in the random case $(\lambda'(i))$ may be taken as reference values that represent the characteristic length with which each leading direction contributes to the shape of a market whose components were correlated at random. These eigenvalues correspond to the characteristic length of the individual (isolated) components of the market. On the other hand, the eigenvalues obtained from actual data $\lambda_t(i)$ represent the characteristic length of each structure emerging from the dynamics of the market, that is, associated to each leading directions of the market space.

# 4 Results and Discussion

Results were computed in relation to actual daily returns data as well as to random data with the same mean and variance.

## 4.1 The S&P500 effective dimensions

The first set of actual data consists in 249 stocks present in S&P500 from July 1973 to March 2003, considering all the surviving firms for the whole period. Part of the ordered eigenvalue distributions obtained from actual data and random data is shown in Fig.1.

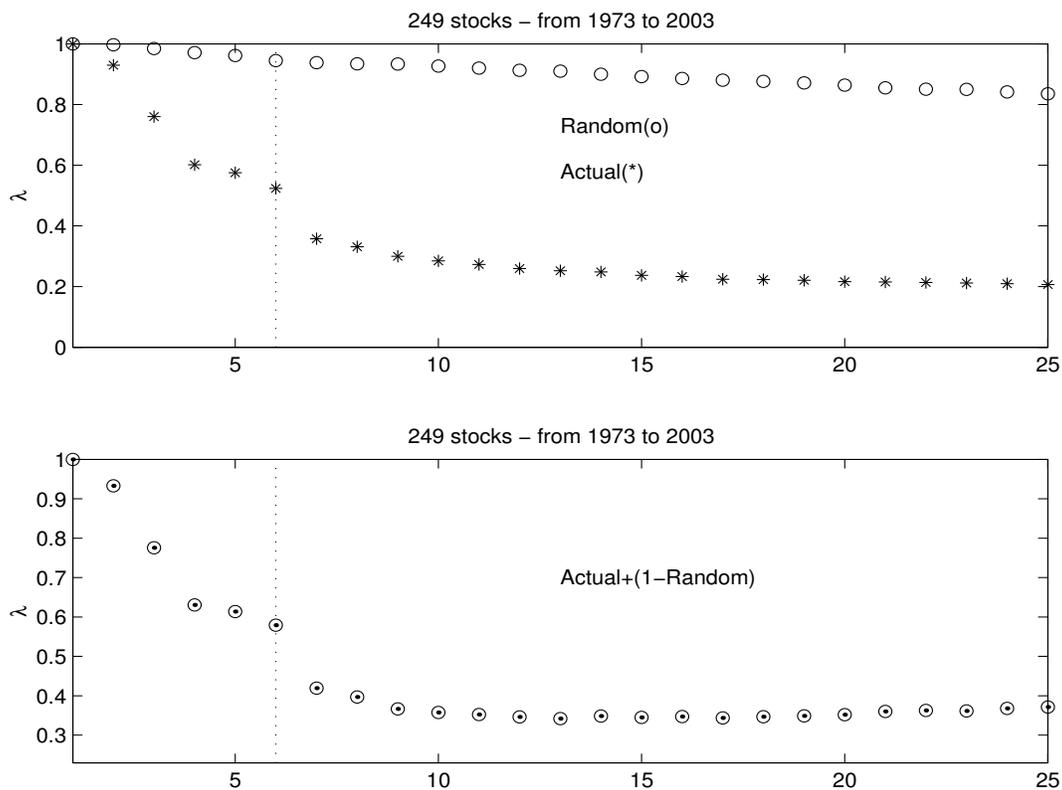

Figure 1: S&P500 249-stocks: decrease of the largest 25 eigenvalues

The plots in Fig.1 represent the largest 25 eigenvalues obtained for the first set of actual data. The largest 25 eigenvalues are compared to the largest 25 eigenvalues obtained from random data. Given the decrease obtained from the $7^{th}$ eigenvalue, we conclude that the market structure is essentially confined to a $6$-dimensional subspace. This proves that this subspace captures the structure of the deterministic correlations that are driving the market and that the remainder of the market space may be considered, for the current purpose, as being generated by random fluctuations.

To test the robustness of this conclusion, we have divided the data in two chronologically successive batches (the first consisting in daily data from July 1973 to March 1988, while the second batch includes data from March 1988 to March 2003) and performed the same operations. In spite of the changes in the market through time, in both cases the behavior of the eigenvalues distribution is very much the same.

Apart from statistical fluctuations, the reconstructed spaces exhibit a reasonable degree of stability, confirming that the number of characteristic dimensions of the S&P500 market space is six. Considering this result, our analysis of the S&P500 market shape is based on 6-dimensional subspaces. The question now is to assess the extent to which the occurrence of extreme phenomena modifies the shape of this subspace and the pattern of behavior of firms and sectors.

## 4.2 The dynamics of crashes

As extreme phenomena are dated events and as we look for their consequences in the distributions of the 6 leading directions, the geometry of the historical data is defined considering short periods. In this sense, instead of the large time intervals that defined the reconstruction of the S&P500 space as in [12], we adopted a 16-days window as the chosen time interval and computed the index of structure with the time window centered at several different dates.

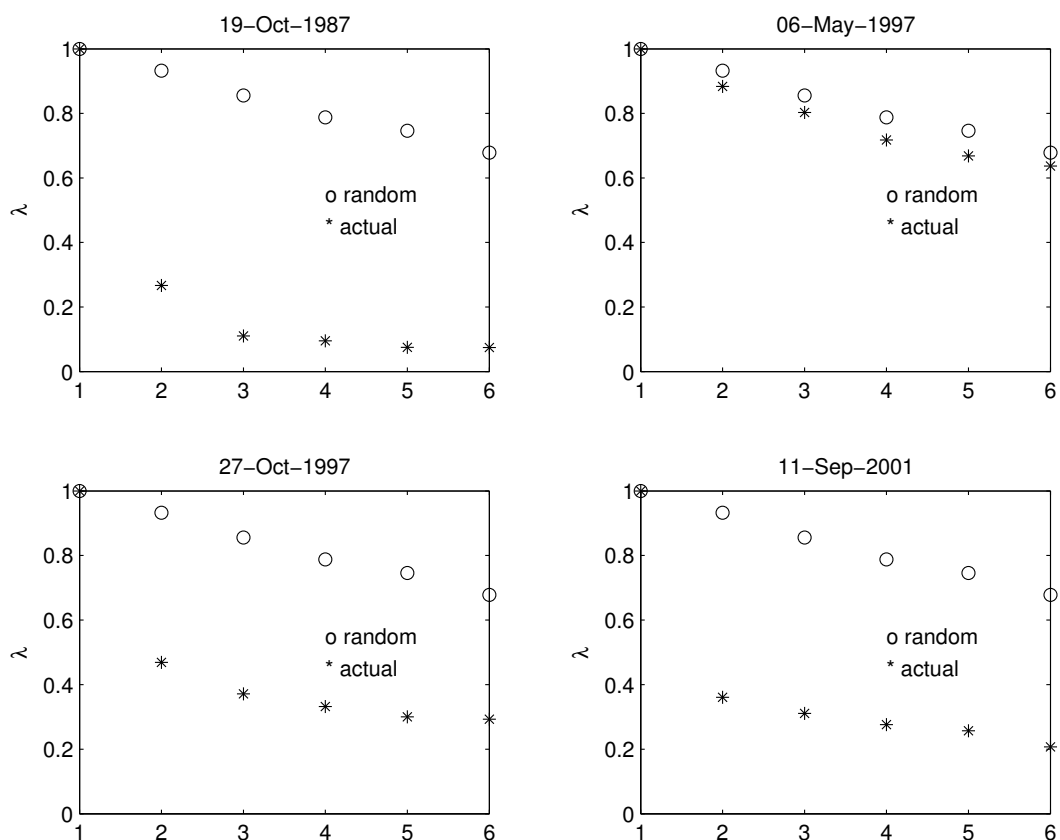

Figure 2: S&P500 deviation from randomness at different dates, comparing crises and a business-as-usual day (6[th] May 1997)

The plots in Figure 2 show some of these dates, namely the crashes of 19[th] October 1987, the *Black Monday*, 11[th] September 2001 and 27[th] October 1997, the *Second Black Monday*. The second plot in this figure shows an unimportant date: May 6, 1997, as suggested in reference [11], was a typical normal day in the US stock market.

The plots in figure 2 show $\lambda(i)$ (with $i=1,\ldots,6$) obtained from the S&P500 market space at four different dates. It is obvious that the values of $S$ obtained for the first and the second *Black Monday*s and for 11[th] September 2001 are high, as there is a great difference in the decrease of the first six eigenvalues computed from actual and random data.

On the contrary, when the same calculation is performed around a typical normal date, the results show that, comparing actual data with random data, there is a quite small difference in the decrease of the first six eigenvalues, which is still another piece of evidence for the robustness of our method.

The geometrical changes in the shape of the market space describe the structural evolution of the characteristic dimensions. As previously indicated, the normal periods qualitatively tend to randomness while the disturbed periods will tend away from randomness.

A less detailed but more extensive result is presented in Fig.3, where the plot shows the daily values of $S$ for the 30 years period. We used a time moving window of 16 days on a market space including the 249 stocks, i.e. all firms surviving through the whole period. The eight highest values of $S$ are marked on the plot.

The highest peaks are identified and correspond to the following crashes:

1. October 1987
2. October 1989
3. October 1997
4. October 1998
5. April 1999
6. Dec.2000/Jan.2001
7. April 2001
8. September 2001

The ranking of the crashes according to the measure of $S$ and its explanation is as follows:

1. October 1987: *Black Monday.*
2. December 2000-Jan.2001: Argentinean Financial crisis (Argentina and Turkey bond market sell-off).
3. October 1989: the US stock market falls almost 7%.
4. September 2001: attack to the Twin Towers.
5. April 1999: Nikkei Crash (Japan).
6. March/April 2001: according to the NBER a recession began in the US in March 2001.
7. October 1998: Russian Crash.
8. October 1997: Asian Crash, the *Second Black Monday.*

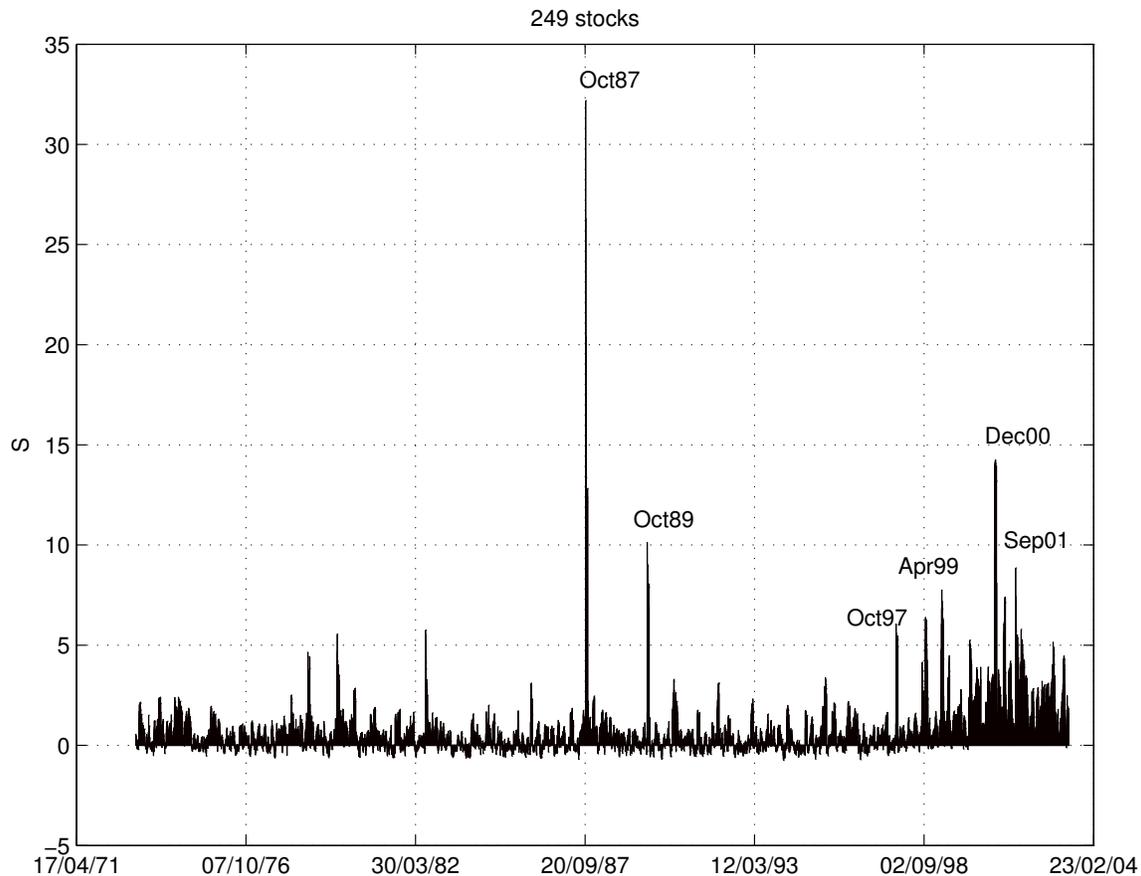

Figure 3: the evolution of the index $S$, measuring the evolution of the S&P500 structure

It is quite obvious from Fig 3 that we have two periods of crises, clustering in 1987-1989 and in 1997-2001: the nature of these periods is discussed below. It should also be considered that some of the events in the list refer to crises in emergent market countries, with considerable effects on the dynamics of the world economy; others refer to the effect of different factors. Indeed, the nature of the triggering factors widely varies. The 1987 crash is well researched and corresponds to a major malfunctioning of the financial system. As Wright points out [21], the Dow Jones suffered a major loss of 22,61% the 19[th] October 1987, whereas the losses were 12,82% the 28[th] October 1929 and 11,73% the 29[th]. Considering the 55 days around the trough, the cumulated loss was of 39,6% in 1929 and of 36,1% in 1987.

Having identified the events corresponding to the eight highest values (peaks) of $S$ in the last 30 years (Fig.3), we reconsidered our data investigating the periods around each peak. Besides providing a more accurate view of the evolution, it allows for a better measurement since at each window we consider a larger number of companies in the S&P500. For the purpose of comparison, the first plot in Fig.4 shows the behavior of $S$ in the nearby of the highest peak compared to the values of $S$ around a typical normal day in the US stock market.

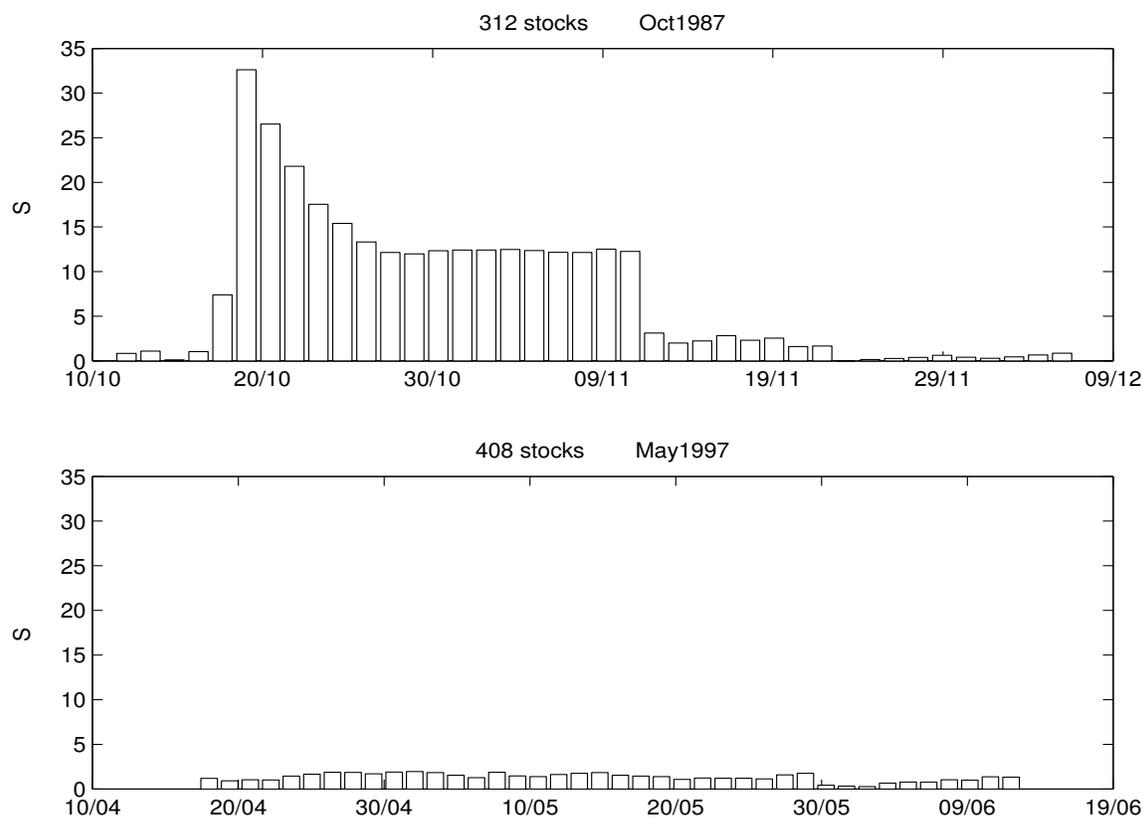

Figure 4: The *Black Monday* and a day of *business-as-usual*

Considering shorter spans of time, we could include larger sets of stocks for each period; consequently, all the entrant firms at each period can be taken into our picture. This procedure highlighted the importance of another crash, which was previously hidden by our selection of the thirty years' survivors. In fact, when the window used for scanning through our data is 40 days, the highest peaks ($S_{Max} = max\{S_t\}\ t_i \leq t \leq t_i+40$ ) organize in the following order (Table 1):

| Ranking | Date (T) | $S_{Max}$ | Number of Stocks included |
|---------|----------|-----------|---------------------------|
| 1 | October 1987 | 31.6 | 312 |
| 2 | Dec.2000/January 2001 | 14.2 | 426 |
| 3 | October 1989 | 10 | 330 |
| 4 | April 2001 | 7.7 | 426 |
| **5** | **April 2000 (NASDAQ)** | **7.6** | **424** |
| 6 | April 1999 | 7.2 | 417 |
| 7 | October 1997 | 6.3 | 408 |
| 8 | October 1998 | 5.8 | 414 |
| 9 | September 2001 | 5.5 | 426 |

Table 1: Ranking of the crises according to the values of $S_{Max}$

Unsurprisingly, the highest peak corresponds to the *Black Monday,* being not only the larger one but also the long-lasting crisis. The most interesting change in the ranking of crashes concerns the appearance of the NASDAQ collapse in April 2000, which was hidden by the fact that some emerging firms in the nineties were not considered in our previous data set since they did not exist for the whole (30 years) period. Yet, when they are considered, the real picture of a turbulent market appears very clearly: it was in the Information Technology and Telecommunication sector that most speculation and stock activity concentrated in the late nineties, during the *Internet bubble*, and the NASDAQ crash marks its end. This crash proves the dimension of this speculative process. The NASDAQ attained its highest peak by early March 2000, and then its all-time highest loss by April (35% of loss in relation to the peak the previous month).

The different crises are compared in the next figures (Figs.5 and 6). They classify in three groups: (a) a *structural* crisis, (b) *general* crises, and (c) *local* crises. Local crises are shorter and less intense (6 to 9 in our ranking), general crises are longer and more intense (2 to 5 in our ranking), whereas a structural crisis (1 in our ranking) is deeper and more prolonged. According to the values obtained by $S_{Max}$, local crises attain maxima of around 6, general crises from 7 to 15, and the structural crisis more than 30 (Fig.4).

A second criterion for the distinction among these types of crises is the rate of decay of the values of $S_{Max}$. For the cases of local crises, these values decrease rather quickly after the peak (plus the 16-days moving window), proving that the structure-generating behavior is short living after the days of the crash.

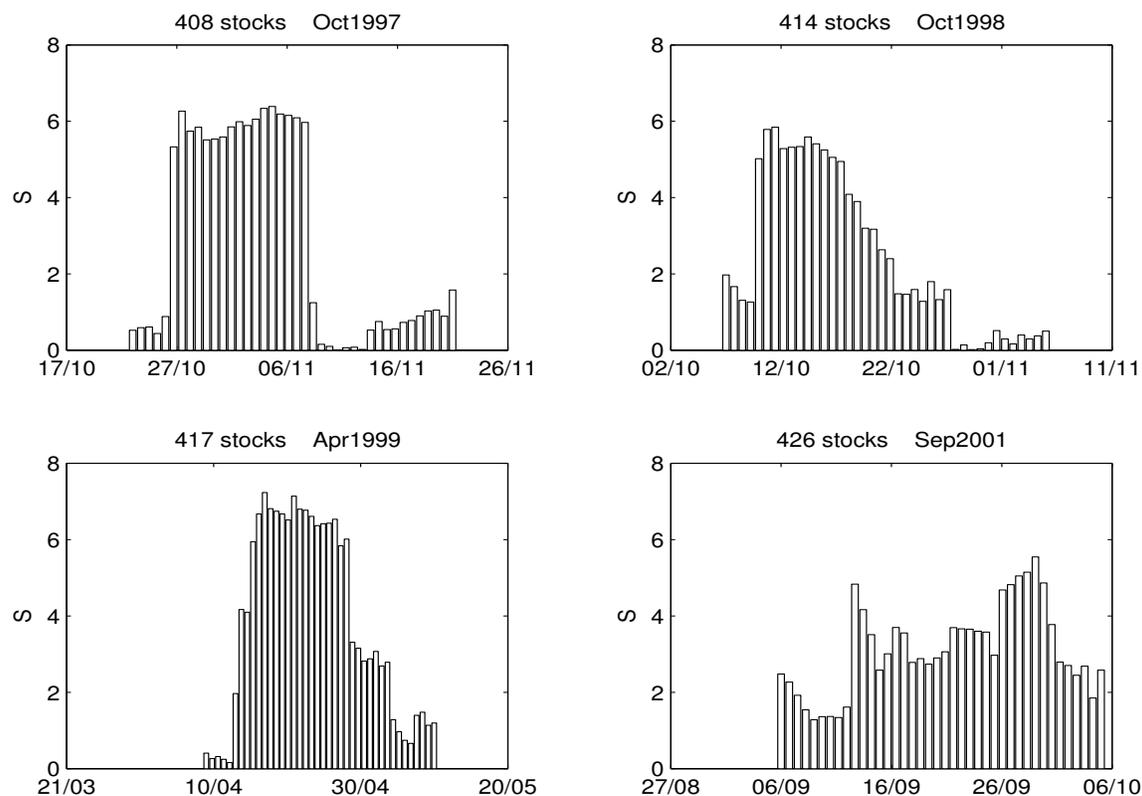

Figure 5: Local crises

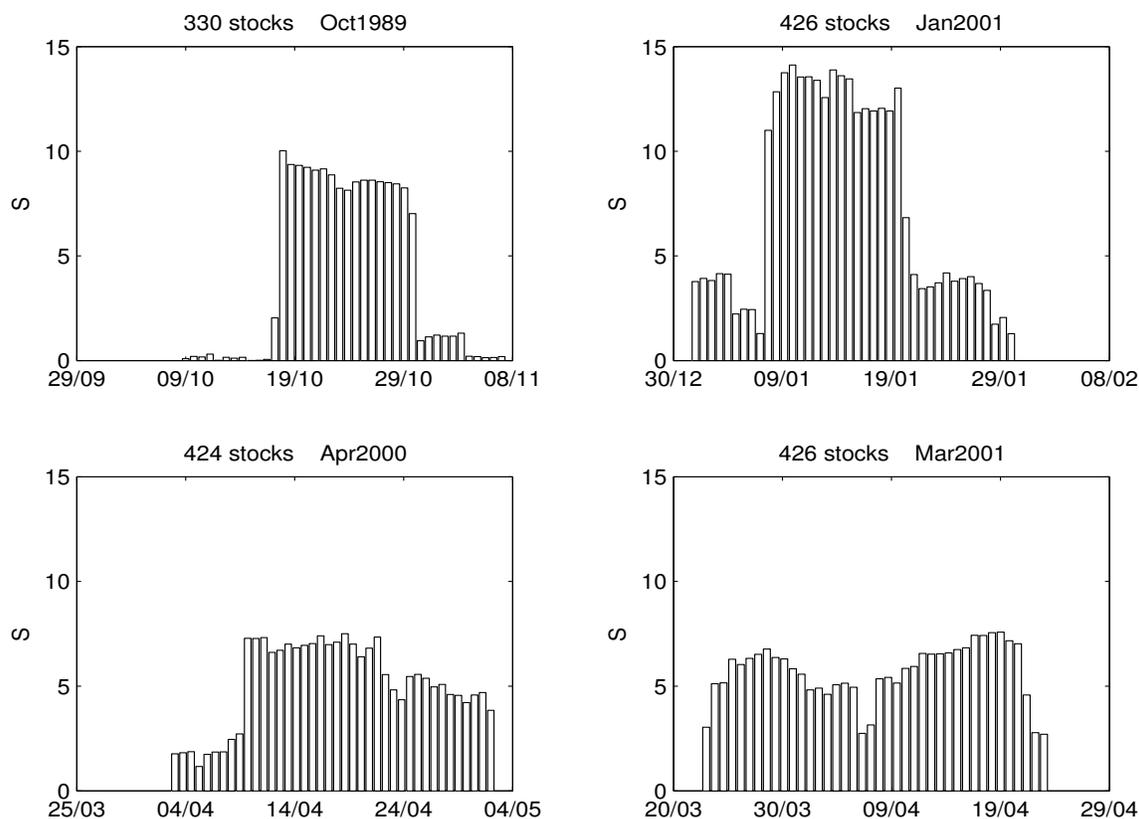

Figure 6: General crises

A third characteristic distinguishing between local and general crises is presented in the next section, where sectoral dynamics is taken into account. Local crises tend to concentrate in some specific sectors; in contrast, general crises tend to exhibit a pattern of perturbation in all sectors (as Fig. 9 shows).

## 4.3 Compared sectoral dynamics

When, instead of the whole set of stocks, we consider sub-sets including the stocks of firms belonging to the same economic sector [1] and compute the index of market structure for each of these sub-sets, evidence for some interesting properties emerges.

In a previous paper and using several topological indexes [12], we verified that in periods of expansion, sector-oriented sub-sets are characterized by a smaller average distance between stocks. The average behavior of companies belonging to the same economic sector is more synchronous than the behavior of the overall market taken as a whole: in the jungle of the crisis, tribes of firms act together. Now we analyze sectoral dynamics by considering the consequences of crashes on the leading directions of nine sector-oriented market spaces, being each of them restricted to stocks in one of the following sectors: Energy, Materials, Industry, Consumer Discretionary, Consumer Staples, Health Care, Financials, Information Technology and Utilities.

---

[1] Detailed structures of sectors and other information from Global Industry Classification Standard (GICS®), available at http://www.standardandpoors.com/, referenced in June, 2005.

In Figs.7 to 9, the histograms show the value of $S_{Max}$ obtained from those nine different market spaces, all of them built on the same time period, which is indicated in the title of the plots. The results show the remarkable impact of the Asian crisis in the Financial sector and the strong effect of the attack to the Twin Towers on the Materials and Industrial sectors.

Again, there is an obvious difference between what we classify as local and as general crises. A third characteristic distinguishing between local and general crises is obvious from the graphs. Local crises tend to concentrate in some specific sectors (financial companies for the Asian Crash, industrial, materials and financial companies for the case of the reaction to the 11[th] September). In contrast, general crises tend to exhibit a pattern of perturbation in all sectors (as Fig. 9 shows).

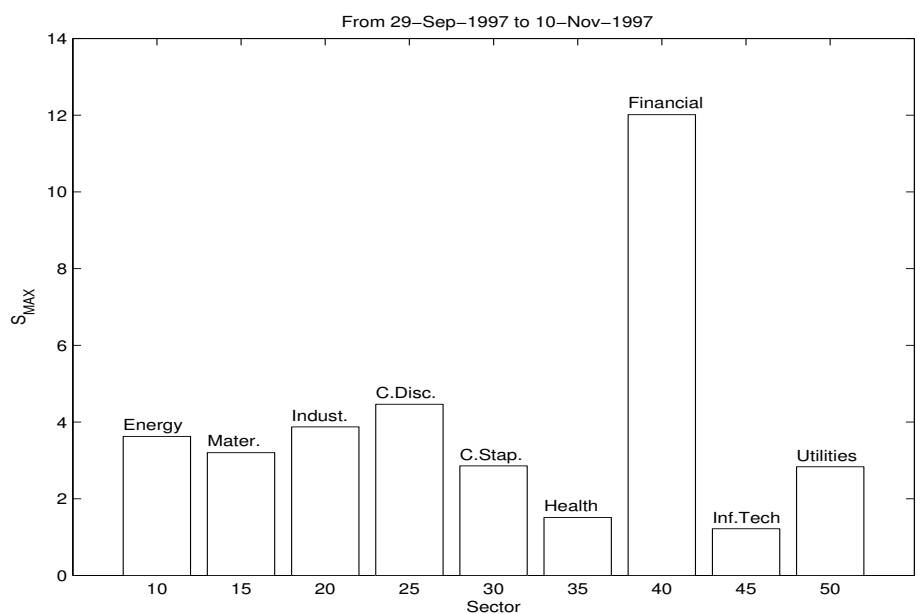

Figure 7: Asian Crash

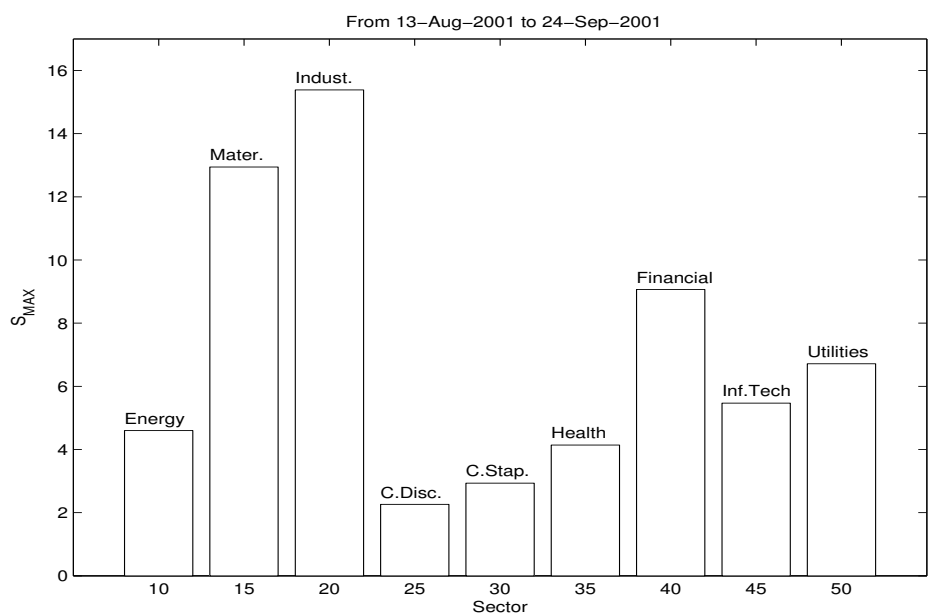

Figure 8: September 11[th]

The plot in Fig.9 shows the extraordinarily unique character of the 1987 *Black Monday*: this is the only case of a crash provoking a similar dynamics in all major sectors, whereas in all other crises the dynamics and time pattern of the main sectors is clearly divergent.

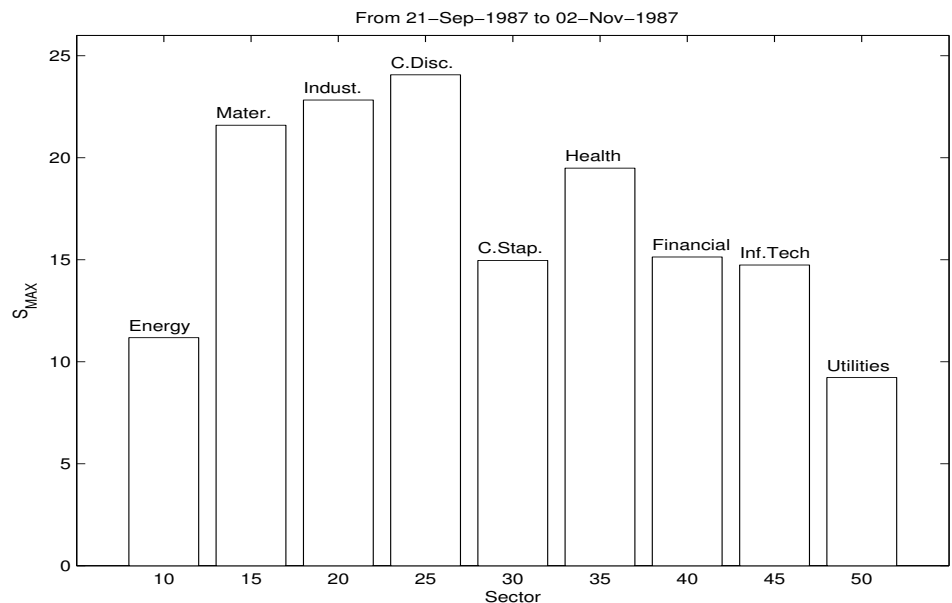

Figure 9: Black Monday

Finally, we compare the sectoral dynamics among different crashes, taking the examples of Materials and Financials. Because in the *Black Monday* crisis the index $S$ reaches very high values in all sectors, this crisis was intentionally excluded from the plots in Fig.10.

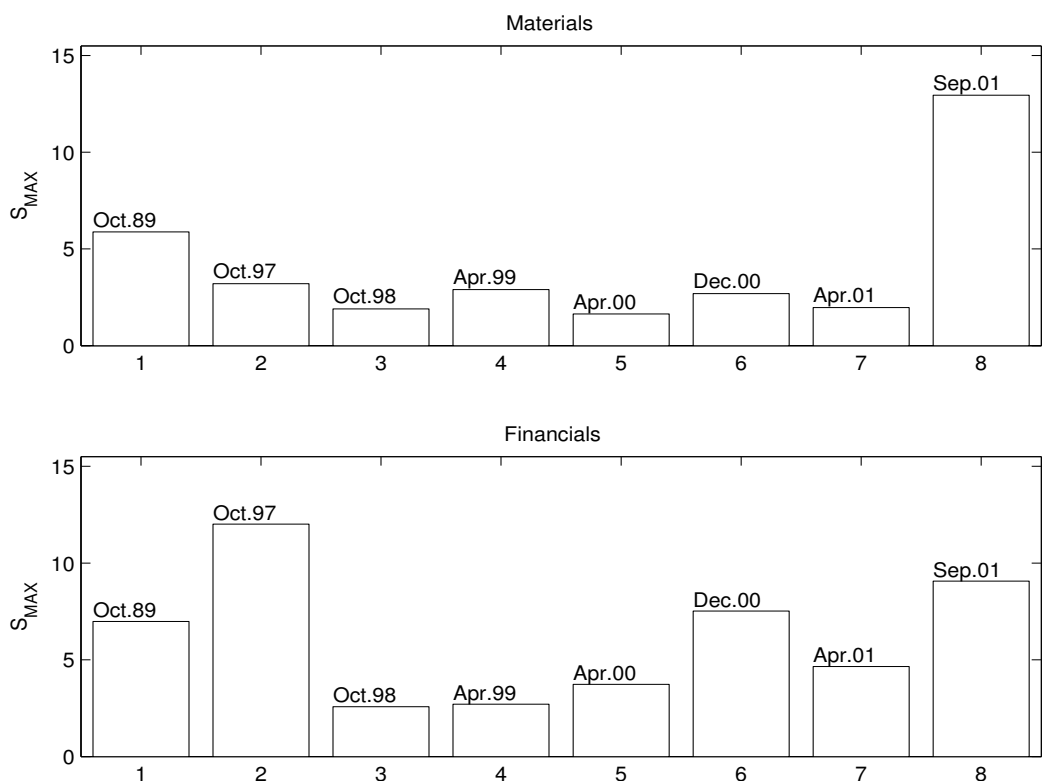

Figure 10: Materials and Financials dynamics

The following table summarizes the sectoral pattern of the crashes, indicating the sectors leading the structural change:

| Date | Leading Sectors |
|---|---|
| October 1987 *Black Monday* | All |
| January 2001 Argentinean Crisis | Financials |
| October 1989 US stock market | Consumer Staples/Financials |
| September 2001 Twin Towers | Industrials/Materials/Financials |
| April 2000 NASDAQ | Information Technology (IT) |
| October 1998 Russian Crash | Energy/Utilities |
| April 1999 Nikkei Crash | Consumer Discretionary |
| April 2001 US recession | Energy/IT |
| October 1997 Asian Crash | Financials |

Table 2: description of the sectors dominating each crash

From the above results, one notices that the Financials sector is the sector that most frequently appears as a leading sector. Its appearance as the leading sector of both the Argentinean and the Asian crises is in accordance with the appropriate expectations, since each of these crises corresponds to a major malfunctioning of the financial system. Another encouraging result refers to the Information Technology leadership at the NASDAQ crisis, settling the end of the Internet Bubble of the second half of the nineties.

Back to the geometrical tale of our index, a 3 dimensional look at the market space that evolves from October 1989 to September 2001 and comprises on average 80 Financial stocks (the lower plot in Fig.10), would reveal a manifold that: *(i)* starts from a elliptical form (in 1989), *(ii)* acquires prominences in a particular direction at the 1997 *Asian Crash*, and *(iii)* turns back to a close-to-spherical form until the Argentinean Financial crisis in December 2000. After a partial shape recovery, a new relevant distortion will arrive in September, 2001.

A smoother dynamics characterizes the market space built from stocks in the Materials sector along the same time period (1989 to 2001). Accordingly to the results presented above (the upper plot in Fig.10), the only relevant shape distortion of that market space is the one taking place in 11$^{th}$ September, 2001; when the structure index $S$ reaches a value three times higher than the highest value obtained so far for the Materials market space.

## 4 Conclusions

A stochastic geometry technique proved to be useful for the purpose of describing and interpreting the evolution and changes in the dynamics of a market. Furthermore, the index $S$, as defined in this paper, allowed for a useful taxonomy of the nine major stock market crises occurring in the last thirty years. Three types of crises were considered: local, general and structural crises. We classified these crashes according to the maximum values of $S$, but three other operative criteria were useful to describe these differences: (1) the decay time of the effects of the crash is reduced in the case of local crises; (2) general crises concentrate on several sectors; (3) the structural crisis involves all sectors under a similar time pattern. The measure of $S_{Max}$ proved to be useful and capable of discriminating among the distinct processes at work in the stock market.

As the index *S* captures the lack of uniformity among the market effective dimensions, we are able to characterize the extent to which crashes act differently on specific directions, causing changes in the shape of the market space. Looking for relevant distortions in the shape of the S&P500 market space through the last 30 years, we identified the events corresponding to crises in emergent market countries, with considerable effects on the dynamics of the world economy. Others events that were also identified refer to the effect of different factors, showing that, the nature of the triggering factors widely varies.

The identification of the characteristics of each crisis allows for their differentiation. Local crises were imposed either by disarrangements of national stock markets from emerging economies (Russia, Asia) and global players (Japan) or by purely exogenous factors (the 11[th] September attack). The crash provoked by exogenous factors is less consequential and is rapidly superseded. Instead, general crises followed another pattern: they are deeper, longer and involve a large number of sectors. The Argentinean crises (December 2000-January 2001) and the following NASDAQ crisis (April 2000) and the US recession (April 2001) initiated or followed the end of the Internet Bubble of the second half of the nineties.

The *Black Monday* (1987) was the deeper and the longest of all the crashes as well as the more general, since it involved all economic sectors. The data suggest that another structural crisis may be at work in the clustering of six crashes between April 1997 and September 2001.

Finally, the predictive character of our structure index is to be explored in future work.

## References

[1] Mantegna, R.(1999)," R. N. Mantegna, *Hierarchical Structure in Financial Markets*, Eur. Phys. J. B **11**, 193-197.
[2] Mantegna, R, Stanley, H., (2000), *An Introduction to Econophysics: Correlations and Complexity in Finance*, Cambridge: Cambridge University Press.
[3] Bonanno, G. *et. al*, (2004), "Networks of Equities in Financial Markets", Eur Phys J B, 38, 363-371.
[4] Bonanno, G. *et. al*, (2004), "Topology of Correlation based Minimal Spanning Trees in Real and Model Markets", eprint arXiv:cond-mat/0211546.
[5] Bonanno, G. *et. al*, (2001), "High-frequency Cross-correlation in a Set of Stocks", Quantitative Finance, 1, 96-104.
[6] Matteo M, *et. al*, "An interest rates Cluster Analysis", Physica A 339 (2004) 181-188.
[7] Kullmann, Kertesz, Mantegna, R. (2000), "Identification of Clusters of Companies in Stock Indices via Potts Super-paramagnetic Transitions", Physica A 287, 412-419.
[8] J.-P. Onnela, A. Chakraborti, K. Kaski, J. Kertesz (2003), "Dynamics of Market Correlations: Taxonomy and Portfolio Analysis", Phys. Rev. E 68, 056110.
[9] P.Gopikrishnan, B. Rosenow, V. Plerou, and Stanley, H., (2001), "Identifying Business Sectors from Stock Price Fluctuations", Phys. Rev. E, 64, 035106R.
[10] Marsili, M., (2002), "Dissecting Financial Markets: Sectors and States", Quantitative Finance, 2, 297-302.
[11] Bonanno, G., Lillo, F., Mantegna, R. (2001), "Levels of Complexity in Financial Markets", Physica A, 299, 16-27.

[12] Vilela Mendes, R.; Araújo, T.; Louçã, F. (2003), "Reconstructing an Economic Space from a Market Metric", Physica A, 323, 635-50.

[13] Vilela Mendes, R., (2001), "Structure Generating Mechanisms in Agent-based Models", Physica A, 295 537-561.

[14] Lillo, F., Mantegna, R. (2002), "Dynamics of a Financial Market Index after a Crash", Physica A 338, 125-134.

[15] J.-P. Onnela, A. Chakraborti, K. Kaski, J. Kertesz (2003), "Dynamic Asset Trees and Black Monday", Physica A 324, 247.

[16] Sornette, D. and Helmstetter,A. (2003) "Endogeneous Versus Exogeneous Shocks in Systems with Memory", Physica A 318, 577.

[17] Sornette, D. (2002): *Why Stock Markets Crash (Critical Events in Complex Financial Systems)*, Princeton University Press.

[18] Sornette, D. (2005), "Endogenous versus Exogenous Origins of Crises", in monograph on extreme events, V. Jentsch editor, Springer.

[19] Sornette, D.(2000), "The Nasdaq Crash of April 2000: Yet another example of log-periodicity in a speculative bubble ending in a crash", European Physical Journal B 17, 319-328.

[20] Sornette, D.(2002), "Endogenous versus Exogenous Crashes in Financial Markets", eprint arXiv:cond-mat/0210509.

[21] Wright, R. (2002): *Chronology of the Stock Market*, McFarland & Company, Inc.